\documentclass{article}
\usepackage{color,ulem,soul}
\usepackage{amsmath, amssymb, amsthm, braket, latexsym, mathtools}
\usepackage{subfiles,subcaption}
\usepackage{bm,array,arydshln}
\usepackage{graphicx}

\newtheorem{theorem}{Theorem}
\newtheorem{lemma}{Lemma}

\newcommand{\C}{\mathbb{C}}

\DeclareRobustCommand{\Erase}{\bgroup\markoverwith{\textcolor{red}{\rule[.5ex]{2pt}{0.4pt}}}\ULon}

\begin{document}
\title{{\bf Sensitivity of Quantum Walk to Phase Reversal and Geometric Perturbations: An Exploration in Complete Graphs}
\vspace{0mm}} 

\author{
  Taisuke Hosaka \\
  \small Graduate School of Environment and Information Sciences \\
  \small Yokohama National University \\
  \small Hodogaya, Yokohama, 240-8501, Japan \\
  \\ 
  Renato Portugal \\
  \small National Laboratory of Scientific Computing (LNCC) \\
  \small Petr\'{o}polis, RJ, 25651-075, Brazil \\
  \\ 
  Etsuo Segawa \\
  \small Graduate School of Environment and Information Sciences \\
  \small Yokohama National University \\
  \small Hodogaya, Yokohama, 240-8501, Japan \\
}

\date{\empty }

\maketitle
\thispagestyle{empty}




\par\noindent


\begin{abstract}
In this paper, we analyze the dynamics of quantum walks on a graph structure resulting from the integration of a main connected graph $G$ and a secondary connected graph $G'$. This composite graph is formed by a disjoint union of $G$ and $G'$, followed by the contraction of a selected pair of vertices creating a cut vertex $v^*$ and  leading to a unique form of geometric perturbation. Our study focuses on instances where $G$ is a complete graph $K_N$ and $G'$ is a star graph $S_m$. The core of our analysis lies in exploring the impact of this geometric perturbation on the success probability of quantum walk-based search algorithms, particularly in an oracle-free context. Despite initial findings suggesting a low probability of locating the perturbed vertex $v^*$, we demonstrate that introducing a phase reversal to the system significantly enhances the success rate. Our results reveal that with an optimal running time and specific parameter conditions, the success probability can be substantially increased. The paper is structured to first define the theoretical framework, followed by the presentation of our main results, detailed proofs, and concluding with a summary of our findings and potential future research directions.

    \
    
\end{abstract}
  

\section{Introduction}

Quantum walks have been explored in the literature as the quantum counterpart of classical random walks. The coined model, one of the most extensively studied discrete-time quantum walk models, was initially analyzed on the line in \cite{ADZ93} and subsequently extended to graphs in \cite{AAKV01}. Its time evolution operator consists of the coin operator and the shift operator. The coin operator alters the internal state of a particle located at a vertex, while the shift operator relocates the particle to its neighboring vertices, conditioned by its internal state~\cite{HKSS14}. The coined quantum walk model has been instrumental in developing quantum search algorithms aimed at locating a target within a graph, representing one of the significant applications of quantum walks. The fundamental concept behind constructing quantum search algorithms involves an oracle that inverts the phase when the particle is on the target, while maintaining the state unaltered for non-target vertices. For instance, \cite{AKR05} discusses a search algorithm that is quadratically faster than classical random walks on a two-dimensional torus. This is achieved by modifying the coin operator to invert the phase when the particle is on the target and applying the Grover operator on unmarked vertices. Furthermore, search algorithms have also been applied to various graphs, such as the hypercube \cite{SKW03} and Johnson graphs \cite{TSP22}.

In this paper, we analyze a quantum walk on a main connected graph $G$ that is modified by a second connected graph $G'$. The modification involves making a disjoint union of $G$ and $G'$, followed by contracting a pair of vertices $v_1$ from $G$ and $v_2$ from $G'$. This contraction means identifying vertices $v_1$ and $v_2$ as a single vertex, denoted $v^*$. The resulting graph is connected, a process we refer to as geometric perturbation located at the cut vertex or articulation point $v^*$. We focus on cases where $G$ is a complete graph $K_N$ and $G'$ is a star graph $S_m$ with $m$ leaves. Here, $v^*$ is the contraction of the center vertex of the star graph with an arbitrary vertex of $K_N$. Our objective is to demonstrate that the geometric perturbation of the original graph can be a method to mark a specific vertex, useful in oracle-free quantum-walk-based search algorithms.

We hypothesize that if a quantum walker is influenced by the geometric perturbation, the probability of finding $v^*$ increases. However, our findings indicate minimal response from the quantum walker to this perturbation. Through numerical simulations, we observe that the probability of finding vertex $v^*$ is approximately $1/N$, which is disappointingly low. To address this, we introduce an additional perturbation to the dynamics by reversing the phase for each leaf of $S_m$, a process we refer to as phase reversal. When combining geometric perturbation and phase reversal, the success probability for locating $v^*$ increases to $1/2 + o(1)$ for any value of $\alpha$, where $\alpha$ is a parameter that determines the number of leaves in $S_m$ through the expression $m = \lfloor N^\alpha \rfloor$.

Furthermore, we determine that the optimal running time is $O(N^{(2-\alpha)/2})$ for $\alpha \in [0, 1]$. This outcome suggests that, while geometric perturbation contributes to the acceleration of the search, the phase reversal significantly boosts the probability of success. Additionally, we establish the optimal time step for cases where $\alpha \geq 1$, which is $O(\sqrt{N})$ regardless of the value of $\alpha$. This result indicates a phase transition at $\alpha = 1$, beyond which the optimal time step does not improve further.

This paper is organized in the following manner: Section 2 presents the formal definitions of our setting. Section 3 outlines the main results summarized in two theorems. In Section 4, we provide the proofs of these theorems. Finally, Section 5 present our final remarks.

\section{Our setting}
Let $G=(V,E)$ be a simple and connected graph with the set $V=V(G)$ of vertices and the set $E=E(G)$ of edges.
In addition, $A=A(G)$ is the set of symmetric arcs induced by $E(G)$.
For $a \in A$, the origin and terminus of $a$ are denoted by $o(a)$ and $t(a)$, respectively.
Additionally, let $\Bar{a}$ be the inverse arc of $a$.
The complete graph with $n$ vertices, denoted by $K_{n}$, is the graph in which any two vertices are adjacent.
The star graph, denoted by $S_{n}$, is the tree with one center vertex and $n$ leaves.
The degree of the vertex $v \in V$ is denoted by $\mathrm{deg}(v)=|\{a \in A : t(a)=v\}|$.
Let $K_{N} \wedge_{v^*} S_{m}$ be the graph identifying a fixed vertex of $K_{N}$ with the center vertex of $S_m$.
This identified vertex is denoted by $v^{*}$.
This definition shows that $v^{*}$ has a geometric perturbation by $S_{m}$ compared to $v \in V_{K} \setminus \{v^{*}\}$.
The sets of the arcs and vertices are decomposed into the following disjoint unions:
\begin{align*}
    &A:=A(K_{N} \wedge_{v^*} S_{m})= A(K_{N}) \sqcup A(S_{m}), \\
    &V:=V(K_{N} \wedge_{v^*} S_{m})= V_{K} \sqcup V_{S} \sqcup \{v^{*}\},
\end{align*}
where 
\begin{align*}
    V_{K}=V_{K} \setminus \{v^{*}\}, \quad V_{S}=V_{S} \setminus \{v^{*}\}.
\end{align*}

The time evolution operator  $U: \C^{A} \rightarrow \C^{A}$  of a quantum walk on $K_{N} \wedge_{v^*} S_{m}$ is defined as
\begin{align*}
    U=S\left(2d^{*}d-I\right),
\end{align*}
where $S$ is the shift operator, $d$ is the boundary operator, and $I$ is the identity operator.
The shift operator on $\mathbb{C}^{A}$ is defined as
\begin{align*}
    (S\psi)(a)=\psi(\bar{a}),
\end{align*}
for any $a \in A$ and $\psi \in \mathbb{C}^{A}$.
The boundary operator $d: \mathbb{C}^{A} \rightarrow \mathbb{C}^{V}$ is defined as
\begin{equation*}
    (d\psi)(v)=
    \begin{dcases}
        \frac{1}{\sqrt{\mathrm{deg}(t(a))}} \sum_{t(a)=v}\psi(a) &: t(a)=v, v \notin V_{S}, \\
        0 &: \mathrm{otherwise}. \\
    \end{dcases}
\end{equation*}
The adjoint operator $d^{*}: \mathbb{C}^{V} \rightarrow \mathbb{C}^{A}$ is given by
\begin{equation*}
    (d^{*}f)(a)=
    \begin{dcases}
        \frac{1}{\sqrt{\mathrm{deg}(t(a))}} f(t(a)) &: t(a)=v, v \notin V_{S}, \\
        0 &: \mathrm{otherwise}. \\
    \end{dcases}
\end{equation*}
Note that $S^{2}=I$ and $dd^{*}=I$.
For $a \in A$ and $\psi \in \mathbb{C}^{A}$, the action of the evolution operator is described by 
\begin{equation}
    \label{eq: U}
    \left(U\psi\right)(a)=
    \begin{dcases}
        -\psi(\Bar{a})+\frac{2}{\mathrm{deg}(t(a))}\sum_{b \in A:\, o(b)=t(a)}\psi(b) &: t(a) \notin V_{S},\\
        -\psi(\bar{a}) &: t(a) \in V_{S}.\\
    \end{dcases}
\end{equation}
This definition shows that each leaf has its phase reversed after the action of $U$.

Let $\psi_{t}$ be the $t$-th iteration of $U$, that is, $\psi_{t+1}=U\psi_{t}$.
The initial state $\psi_{0} \in \mathbb{C}^{A}$ is the uniform  state on $K_{N}$ expressed as 
\begin{align*}
    \psi_{0}(a)=
    \left\{
    \begin{array}{ll}
          {\frac{1}{\sqrt{N(N-1)}}}  &: a \in A(K_{N}), \\
          0                        &: \mathrm{otherwise}.
    \end{array}
    \right.
\end{align*}
The probability of finding the vertex $v$ at time $t$ is obtained from
\begin{align*}
    p_{t}(v)= \sum_{a \in A : t(a)=v} |\psi_{t}(a)|^{2}.
\end{align*}
We are particularly interested in finding the optimal running time $t_\text{opt}$, at which the probability $p_{t_\mathrm{opt}}(v^*)$ of finding vertex $v^*$ is maximized.

\section{Main results}

\begin{figure}[htbp]
    \begin{center}
        \includegraphics[scale=0.5]{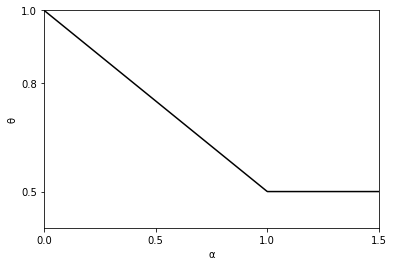}
        \caption{The phase transition diagram for $\theta$ vs $\alpha$, where $\theta$ is defined so that $t_\text{opt}=O(N^{\theta})$.}
        \label{fig:diagram}
    \end{center}
\end{figure}

\begin{figure}[htbp]
    \begin{center}
        \includegraphics[scale=0.6]{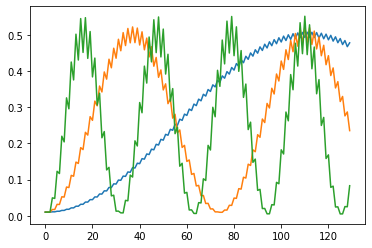}
        \caption{ The time evolution of the probability of finding the marked vertex for $N=100$. The blue and orange curves correspond to the cases for $\alpha=0$ and $\alpha=1/2$, respectively, while the green curve corresponds to the case for $\alpha=1$.}
        \label{fig:prob}
    \end{center}
\end{figure}

In this section, we present our findings on the probability of detecting a perturbed vertex and the optimal running time.
\begin{theorem}
    \label{thm:p}
    For a sufficiently large $N$ and any $\alpha \in [0,\infty)$, there exists a time $t_\text{opt}$ such that 
    \begin{align*}
        p_{t_\mathrm{opt}}(v^{*}) = \frac{1}{2}+o(1).
    \end{align*}
\end{theorem}
Theorem~\ref{thm:p} indicates that the probability is approximately $1/2$, independent of the intensity of the geometric perturbation $\alpha$. Therefore, it is the phase reversal, not the geometric perturbation, that ensures a high probability of detecting the perturbed vertex. The subsequent question might concern the contribution of the parameter $\alpha$. This is addressed in Theorem~\ref{thm:t}. 
\begin{theorem}
    \label{thm:t}
    For a sufficiently large $N$, the optimal running time $t_\mathrm{opt}$ is 
    \begin{align*}
        t_\mathrm{opt}=
        \left\{
            \begin{array}{ll}
                O(N^{\frac{2-\alpha}{2}}) &: 0 \leq \alpha \leq 1, \\
                O(\sqrt{N}) &: 1 < \alpha.
            \end{array}
        \right.
    \end{align*}
\end{theorem}
Theorem~\ref{thm:t} implies that as $\alpha$ increases, the efficiency of the optimal time step improves until $\alpha \leq 1$. Beyond that point, once $\alpha$ exceeds $1$, the optimal time step remains at $O(\sqrt{N})$ and does not enhance further. This implies that the intensity of the geometric perturbation plays a role in accelerating the search process. Fig.~\ref{fig:diagram} illustrates the relationship between the behavior of the exponent of the optimal time and the parameter $\alpha$. It shows a phase transition in the search speed occurring at $\alpha=1$. Additionally, Fig.~\ref{fig:prob} displays a numerical simulation of the success probability for various values of $\alpha$.

\section{Proof of main results}
First, $A$ is divided into the five disjoint sets as follows:
\begin{align*}
    A_{0}&=\{a \in A \,|\, o(a), t(a) \in V_{K}\}, \\
    A_{K}^{+}&=\{a \in A \,|\, o(a) \in V_{K}, t(a)=v^{*} \}, \\
    A_{K}^{-}&=\{a \in A \,|\, o(a)=v^{*}, t(a) \in V_{K} \}, \\
    A_{S}^{+}&=\{a \in A \,|\, o(a) \in V_{S}, t(a)=v^{*} \}, \\
    A_{S}^{-}&=\{a \in A \,|\, o(a)=v^{*}, t(a) \in V_{S} \}. \\
\end{align*}
Note that the size of each arc set is
\begin{align*}
    |A_{0}|=(N-1)(N-2), \quad |A_{K}^{+}|=|A_{K}^{-}|=N-1, \quad |A_{S}^{+}|=|A_{S}^{-}|=m=N^{\alpha}.
\end{align*}
Let us set $\mathcal{A}_{0}=\{A_{0}, A_{K}^{+}, A_{K}^{-}, A_{S}^{+}, A_{S}^{-}\}$ and $\mathcal{V}_{0}=\{V_{K}, V_{S}, \{v^{*}\}\}$.
Then, we define two operators $\mathcal{U}: \mathbb{C}^{A} \rightarrow \mathbb{C}^{\mathcal{A}_{0}}$ and $\mathcal{W}: \mathbb{C}^{V} \rightarrow \mathbb{C}^{\mathcal{V}_{0}}$ such that 
\begin{align*}
    (\mathcal{U}\psi)(B)=\frac{1}{\sqrt{|B|}}\sum_{b \in B}\psi(b), \quad 
    (\mathcal{W}f)(W)=\frac{1}{\sqrt{|W|}}\sum_{v \in W}f(v),
\end{align*}
where $B \in \mathcal{A}_{0}$ and $W \in \mathcal{V}_{0}$.
It is easily checked that 
\begin{align*}
    (\mathcal{U}^{*}\Psi)(a)=\frac{1}{\sqrt{|B_{a}|}}\Psi(B_{a}),
\end{align*}
where $B_{a} \in \mathcal{A}_{0}$ which $a \in B_{a}$ holds.
We should remark that $\mathcal{U}\mathcal{U}^{*}$ is the identity operator on $\mathbb{C}^{\mathcal{A}_{0}}$ while $\mathcal{U}^{*}\mathcal{U}=:\Pi_{\mathcal{A}_{0}}$ is a projection onto $\mathbb{C}^{\mathcal{A}_{0}} \subset \mathbb{C}^{A}$.
By definition of $\Pi_{\mathcal{A}_{0}}$, we obtain
\begin{align*}
    (\Pi_{\mathcal{A}_{0}}\psi)(a)=\frac{1}{|B_{a}|}\sum_{b \in B_{a}}\psi(b).
\end{align*}
From the symmetry of $U$, we easily obtain following.
\begin{lemma}
    \label{lem:inv}
    For the projection $\Pi_{\mathcal{A}_{0}}$, we have
    \begin{align*}
        U\Pi_{\mathcal{A}_{0}}=\Pi_{\mathcal{A}_{0}}U.
    \end{align*}
\end{lemma}

\begin{lemma}
    The time evolution operator $U_{0}: \C^{\mathcal{A}_{0}} \rightarrow \C^{\mathcal{A}_{0}}$ is expressed as 
    \begin{align*}
        U_{0}=S_{0}(2 \partial^{*}\partial-I),
    \end{align*}
    where $S_{0}=\mathcal{U}S\mathcal{U}^{*}$ and $\partial=\mathcal{W}d\mathcal{U}^{*}$.
\end{lemma}
We define $T=\partial S_{0} \partial^{*}: \C^{\mathcal{V}_{0}} \rightarrow \C^{\mathcal{V}_{0}}$.
Then the spectrum of $T$ is easily given by
\begin{align*}
    \mathrm{Spec}\left( T \right)=\{\cos \theta_{1}, \cos \theta_{2}\},
\end{align*}
where
\begin{align}
    \label{eq:cos}
    \cos \theta_{x}=\frac{1}{2(N-1)}\left((N-2) +(-1)^{x-1} \sqrt{N^{2}-4N^{\alpha}+\frac{4N^{2\alpha}}{N+N^{\alpha}-1}}\right).
\end{align}
The eigenvector of $T$ associated to $\cos \theta_{x}$ is given by
\begin{align}
    \label{eq:vec_T}
    f_{x}(\mathcal{V}_{0})=\alpha_{x}^{-1}\times
    \left\{ 
        \begin{array}{ll}
        \displaystyle \cos \theta_{x}  &:\mathcal{V}_{0} =V_{K} \\
        \displaystyle \frac{1}{\sqrt{N+N^{\alpha}-1}}   &:\mathcal{V}_{0}=v^{*}
        \end{array}
    \right.
\end{align}
where $\alpha_{x}$ is the normalization constant.

\begin{lemma}
    \label{lem:spec_U}
    The spectrum of $U$ is given by
    \begin{align*}
        \mathrm{Spec}(U)=\{e^{i \theta_{1}}, e^{ -i \theta_{1}}, e^{ i \theta_{2}}, e^{- i \theta_{2}}, -1\}.
    \end{align*}
    Let $\varphi_{\pm \theta_{x}}$ be the eigenvector of $U$ associated to $e^{\pm i \theta_{x}}$. Then we have
    \begin{equation}
        \label{eq:vec_theta}
        \varphi_{\pm \theta_{x}}(\mathcal{A}_{0})=\frac{1}{\sqrt{2}\alpha_{x}|\sin \theta_{\lambda}|} \times 
        \begin{dcases}
              \sqrt{\frac{N-2}{N-1}} \cos \theta_{x} (1-e^{\pm i \theta_{x}}) &: \mathcal{A}_{0}=A_{0} \\
             \frac{1}{\sqrt{N-1}}\left(\cos \theta_{x}-e^{\pm i \theta_{x}}\cdot \frac{N-1}{N+N^{\alpha}-1} \right) &: \mathcal{A}_{0}=A_{K}^{+} \\
             \frac{1}{\sqrt{N-1}}\left(\frac{N-1}{N+N^{\alpha}-1}-\cos \theta_{x} \cdot e^{\pm i \theta_{x}}  \right) &: \mathcal{A}_{0}=A_{K}^{-} \\
             -\frac{e^{\pm i \theta_{x}}N^{\alpha/2}}{N+N^{\alpha}-1} &: \mathcal{A}_{0}=A_{S}^{+} \\
             \frac{N^{\alpha/2}}{N+N^{\alpha}-1} &: \mathcal{A}_{0}=A_{S}^{-} \\
        \end{dcases}
    \end{equation}
    The eigenvector of $U$ associated to $-1$ is expressed as 
    \begin{equation}
        \label{eq:vec_-1}
        \varphi_{-1}(\mathcal{A}_{0})=\beta^{-1} \times
        \begin{dcases}
              -1 &: \mathcal{A}_{0}=A_{0} \\
              \sqrt{N-2} &: \mathcal{A}_{0}=A_{K}^{\pm}  \\
             -\sqrt{\frac{(N-1)(N-2)}{N^{\alpha}}} &: \mathcal{A}_{0}=A_{S}^{\pm} \\
        \end{dcases}
    \end{equation}
    where $\beta$ is the normalization constant.
\end{lemma}

By using Eqs.~(\ref{eq:vec_theta}) and~(\ref{eq:vec_-1}), we obtain the explicit expression for $\Psi_{t}(\mathcal{A}_{0})=U^{t}\Psi_{0}(\mathcal{A}_{0})$.
Note that the probability of finding $v^{*}$ is given by
\begin{align*}
    p_{t}(v^{*})=|\Psi_{t}(A_{K}^{+})|^{2}+|\Psi_{t}(A_{S}^{+})|^{2}.
\end{align*}
Therefore, we want to know only the expressions of $\Psi_{t}(A_{K}^{+})$ and $\Psi_{t}(A_{K}^{+})$.
Then, we obtain
\begin{equation}
    \label{eq:Psi_t}
    \begin{aligned}
        \Psi_{t}(A_{K}^{+})&=\sum_{x=\{1,2\}} c_{x}\cdot k_{x}+(-1)^{t}\cdot R_{K}, \\
        \Psi_{t}(A_{S}^{+})&=-\sum_{x=\{1,2\}} c_{x}\cdot s_{x}+(-1)^{t+1}\cdot R_{S},
    \end{aligned}
\end{equation}
where 
\begin{align*}
    c_{x}&=\frac{2 \sin \left(\frac{\theta_{x}}{2}\right) }{\alpha_{x}^{2}\sin^{2}{\theta_{x}}\sqrt{N}}\left(\cos \theta_{x}+\frac{1}{N+N^{\alpha}-1}\right), \\
    k_{x}&=\cos\theta_{x}\cdot \sin\left(t\theta_{x}+\frac{\theta_{x}}{2}\right)-\frac{N-1}{N+N^{\alpha}-1}\cdot\sin\left(t\theta_{x}+\frac{3\theta_{x}}{2}\right), \\
    s_{x}&=\frac{\sqrt{N^{\alpha}(N-1)}}{N+N^{\alpha}-1}\cdot\sin\left(t\theta_{x}+\frac{3\theta_{x}}{2}\right), \\
    R_{K}&=\frac{N-2}{\beta^{2}\sqrt{N}}, \\
    R_{S}&=\frac{N-2}{\beta^{2}}\sqrt{\frac{N-1}{N^{\alpha+1}}}.
\end{align*}
Let us analyze the asymptotic behavior of $|\Psi_{t}(A_{K}^{+})|$ and $|\Psi_{t}(A_{S}^{+})|$ for a sufficiently large $N$.
By Eq.~(\ref{eq:cos}), $\cos \theta_{1}$ is estimated by
\begin{equation}
    \label{eq:approx_cos}
    \cos \theta_{1}\approx
    \begin{dcases}
        1-N^{\alpha-2} &: 0 \leq \alpha <1, \\
        1-\frac{1}{2}N^{-1} &: \alpha=1, \\
        1-N^{-1} &: \alpha >1.
    \end{dcases}
\end{equation}
Since $\cos \theta_{1}$ tends to $1$ as $N \rightarrow \infty$,
$\theta_{1}$ is approximated as $\sin \theta_{1}$.
Thus we get
\begin{equation}
    \label{eq:approx_sin}
    \theta_{1} \approx \sin \theta_{1}= \sqrt{1-\cos^{2} \theta_{1}} \approx
    \begin{dcases}
        \sqrt{2}N^{\frac{\alpha-2}{2}} &: 0 \leq \alpha <1, \\
        N^{-\frac{1}{2}} &: \alpha=1, \\
        \sqrt{2}N^{-\frac{1}{2}} &: \alpha >1.
    \end{dcases}
\end{equation}
By using Eqs.~(\ref{eq:approx_cos}) and~(\ref{eq:approx_sin}), we obtain the estimation of $c_{x}, k_{x}, s_{x}, R_{K}$ and $R_{S}$.
\begin{lemma}
    \label{lem:cks}
    For above $c_{x}, k_{x}, s_{x}, R_{S}$ and $R_{K}$, these are estimated as follows:
    \begin{equation*}
        c_{1}=
        \begin{dcases}
                \frac{1}{\sqrt{2}}N^{\frac{1-\alpha}{2}}+o(N^{\frac{1-\alpha}{2}}) &: 0 \leq \alpha <1, \\
                1+o(1) &: \alpha=1, \\
                \frac{1}{\sqrt{2}}+o(1) &: \alpha>1.
        \end{dcases} 
    \end{equation*}
    \begin{equation*}
        k_{1}=
        \begin{dcases}
            N^{\alpha-1}\sin(t \theta_{1})+o(N^{\alpha-1}) &: 0 \leq \alpha <1, \\
            \frac{1}{2}\sin(t \theta_{1})+o(1) &: \alpha=1, \\
            \sin(t \theta_{1})+o(1) &: \alpha>1.
        \end{dcases}
    \end{equation*}
    \begin{equation*}
        s_{1}=
        \begin{dcases}
            N^{\frac{\alpha-1}{2}}\sin(t \theta_{1})+o(N^{\frac{\alpha-1}{2}}) &: 0 \leq \alpha <1, \\
            \frac{1}{2}\sin(t \theta_{1})+o(1) &: \alpha=1, \\
            o(1) &: \alpha>1.
        \end{dcases} 
    \end{equation*}
    \begin{align*}
        c_{2}k_{2}=o(1), \quad c_{2}s_{2}=o(1), \quad R_{K}=o(1), \quad R_{S}=o(1).
    \end{align*}
    \begin{proof}
        See Appendix A.
    \end{proof}
\end{lemma}

For $0 \leq \alpha <1$, combining Lemma~\ref{lem:cks} and Eq.~(\ref{eq:Psi_t}), we have
\begin{align*}
    |\Psi_{t}(A_{K}^{+})|&\approx\left|\left(\frac{1}{\sqrt{2}}N^{\frac{1-\alpha}{2}}+o(N^{\frac{1-\alpha}{2}})\right)N^{\alpha-1}\sin(t \theta_{1})+o(1)\right| \\
    &=\left|o(1)\right|, \\
    |\Psi_{t}(A_{S}^{+})|&\approx \left|\left(\frac{1}{\sqrt{2}}N^{\frac{1-\alpha}{2}}+o(N^{\frac{1-\alpha}{2}})\right)N^{\frac{\alpha-1}{2}}\sin(t \theta_{1})+o(1)\right| \\
    &=\left|\frac{1}{\sqrt{2}}\sin(t \theta_{1})+o(1)\right|. \\
\end{align*}
For $\alpha=1$, by using similar method, we obtain
\begin{align*}
    |\Psi_{t}(A_{K}^{+})|\approx\left|\frac{1}{2}\sin(t \theta_{1})+o(1)\right|, \quad
    |\Psi_{t}(A_{S}^{+})|\approx\left|\frac{1}{2}\sin(t \theta_{1})+o(1)\right|.
\end{align*}
For $\alpha>1$, we see
\begin{align*}
    |\Psi_{t}(A_{K}^{+})|\approx\left|\frac{1}{\sqrt{2}}\sin(t \theta_{1})+o(1)\right|, \quad
    |\Psi_{t}(A_{S}^{+})|\approx\left|o(1)\right|.
\end{align*}
Thus the success probability is given by
\begin{align}
    \label{eq:p_t}
    p_{t}=\frac{1}{2}\sin^{2}(t \theta_{1})+o(1)
\end{align}
for any $\alpha \in [0, \infty)$.
Therefore, we obtain desired result for Theorem~\ref{thm:p}.
By the representation of Eq.~(\ref{eq:p_t}), the optimal time step is described by
\begin{align*}
    t_\text{opt}=\left\lfloor \frac{\pi}{2\theta_{1}} \right\rfloor
\end{align*}
From Eq.~(\ref{eq:approx_sin}), $t_\text{opt}$ is expressed as
\begin{equation}
    t_\text{opt}=
    \begin{dcases}
        \left\lfloor \frac{\pi}{2\sqrt{2}}N^{\frac{2-\alpha}{2}} \right\rfloor &: 0 \leq \alpha < 1, \\
        \left\lfloor \frac{\pi}{2}N^{\frac{1}{2}} \right\rfloor &: \alpha=1 \\
        \left\lfloor \frac{\pi}{2\sqrt{2}}N^{\frac{1}{2}} \right\rfloor &: \alpha>1 \\
    \end{dcases}
\end{equation}
This completes the proof of Theorem~\ref{thm:t}.

\section{Final remarks}

In this paper, we have analyzed how a quantum walker perceives graph deformation. We considered a discrete-time quantum walk on $K_{N} \wedge_{v^*} S_{m}$, where $N \gg m$. Here, $K_{N} \wedge_{v^*} S_{m}$ represents the graph formed by the disjoint union of the complete graph $K_N$ and the star graph $S_m$ with $m$ leaves, followed by contracting a pair of vertices $v_1\in K_N$ and $v_2\in S_m$, with $v_2$ being the center vertex of $S_{m}$. The identified vertex is labeled $v^*$. Our focus is on the asymptotic behavior of the success probability at the geometrically perturbed vertex $v^*$ for large $N$, where $m=\left\lfloor N^{\alpha} \right\rfloor$ (for $0 \leq \alpha < 1)$.

Initially, we selected the Grover walk as the time evolution model on this graph. However, we observed that there was almost no difference between the cases for $K_{N}$ and $K_{N} \wedge_{v^*} S_{m}$; the success probability for both is close to $1/N$. In view of this, we introduced a phase reversal on each leaf of $K_{N} \wedge_{v^*} S_{m}$ into the dynamics. We have shown that the probability of finding the vertex $v^*$ is $1/2 + o(1)$, with the optimal running time being $O(N^{(2-\alpha)/2})$ for $0 \leq \alpha < 1$. This result suggests that the phase reversal significantly enhances the success probability at the perturbed location, while the intensity of the geometric perturbation accelerates the search speed.

Moreover, we also consider the case for $\alpha \geq 1$ and rigorously derive the phase diagram of the search speed versus the parameter $\alpha \geq 0$. This diagram indicates a phase transition at $\alpha=1$, implying that the effect of this geometric perturbation on speeding up the search is bounded.

Exploring the behavior of success probability on other graphs, such as the two-dimensional torus and the hypercube, is one of the potential directions for future research.

\section*{Acknowledgments}

R.P. was supported by FAPERJ grant number CNE E-26/200.954/2022, and CNPq grant numbers 308923/2019-7 and 409552/2022-4. 
E.S. acknowledges financial supports from the Grant-in-Aid of
Scientific Research (C) Japan Society for the Promotion of Science (Grant No.~19K03616). The authors have no competing interests to declare that are relevant to the content of this article. All data generated or analyzed during this study are included in this published article.

\section*{Appendix A: Proof of Lemma~\ref{lem:cks}}
{\bf Proof.} \\
{\bf (i) (Estimation of $c_{1}$):}
Since $\cos \theta_{1}$ tends to $1$ as $N \rightarrow \infty$,
$\sin \theta_{1}$ and $\sin (\theta_{1}/2)$ are approximated as $\theta_{1}$ and $\theta_{1}/2$, respectively.
Then $c_{1}$ is given by
\begin{align*}
    c_{1}\approx \frac{\cos\theta_{1}+(N+N^{\alpha}-1)^{-1}}{\alpha_{1}^{2}\theta_{1}\sqrt{N}}.
\end{align*}
Note that $\cos \theta_{1}=1+o(1)$ holds, we have
\begin{equation}
    \label{eq:alpha_1}
    \alpha_{1}^{2}=\cos^{2}\theta_{1}+\frac{1}{N+N^{\alpha}-1}=1+o(1),
\end{equation}
for any $\alpha$.
By using Eqs.~(\ref{eq:approx_sin}) and~(\ref{eq:alpha_1}), $c_{1}$ can be approximated.
For $0 \leq \alpha <1$, we get
\begin{align*}
    c_{1} &= \frac{1+o(1)}{(1+o(1))\sqrt{2}N^{(\alpha-2)/2} N^{1/2}} \\
    &=\frac{1}{\sqrt{2}}N^{\frac{1-\alpha}{2}}+o(N^{\frac{1-\alpha}{2}}).
\end{align*}
For $\alpha=1$, we see
\begin{align*}
    c_{1}&=\frac{1+o(1)}{(1+o(1))\sqrt{2}N^{-1/2} N^{1/2}} \\
    &=1+o(1).
\end{align*}
For $\alpha>1$, we have
\begin{align*}
    c_{1}&=\frac{1+o(1)}{(1+o(1))\sqrt{2}N^{-1/2} N^{1/2}} \\
    &=\frac{1}{\sqrt{2}}+o(1).
\end{align*}
{\bf (ii) (Estimation of $k_{1}$):}
For large $N$, $\theta_{1}$ is approximated as sufficiently small.
Thus, we have
\begin{align}
    \label{eq:sin}
    \sin\left(t\theta_{1}+\frac{\theta_{1}}{2}\right)=\sin (t \theta_{1})+o(1), \quad
    \sin\left(t\theta_{1}+\frac{3\theta_{1}}{2}\right)=\sin (t \theta_{1})+o(1).
\end{align}
While we can easily compute
\begin{equation}
    \label{eq:n/nn}
    \frac{N-1}{N+N^{\alpha}-1} =
    \begin{dcases}
        1-N^{\alpha-1}+o(N^{\alpha-1}) &: 0 \leq \alpha <1, \\
        \frac{1}{2}+o(1) &: \alpha=1, \\
        o(1) &: \alpha>1.
    \end{dcases}
\end{equation}
Combining Eqs.~(\ref{eq:approx_cos}),(\ref{eq:sin}) and~(\ref{eq:n/nn}), we get the desired result. \\
{\bf (iii) (Estimation of $s_{1}$):}
By briefly calculation, we see
\begin{equation}
    \label{eq:nn/n}
    \frac{\sqrt{N^{\alpha}(N-1)}}{N+N^{\alpha}-1} =
    \begin{dcases}
        N^{\frac{\alpha-1}{2}}+o(N^{\frac{\alpha-1}{2}}) &: 0 \leq \alpha <1, \\
        \frac{1}{2}+o(1) &: \alpha=1, \\
        o(1) &: \alpha>1.
    \end{dcases}
\end{equation}
Combining Eqs.~(\ref{eq:sin}) with~(\ref{eq:nn/n}) implies
\begin{equation*}
    s_{1}=
    \begin{dcases}
        N^{\frac{\alpha-1}{2}}\sin(t \theta_{1})+o(N^{\frac{\alpha-1}{2}}) &: 0 \leq \alpha <1, \\
        \frac{1}{2}\sin(t \theta_{1})+o(1) &: \alpha=1, \\
        o(1) &: \alpha>1.
    \end{dcases} 
\end{equation*}
{\bf (iv) (Estimation of $c_{2}k_{2}$ and $c_{2}s_{2}$):}
For large $N$, $\cos \theta_{2}$ approximated as 0 by Eq.~(\ref{eq:cos}).
Thus $\sin \theta_{2}$ and $\sin (\theta_{2}/2)$ are approximated as $1$ for large $N$.
Then $\alpha_{2}$ is given by
\begin{align*}
    \alpha_{2}^{2} \approx \frac{1}{N+N^{\alpha}-1}.
\end{align*}
Therefore we get 
\begin{align*}
    c_{2} &\approx \frac{2}{\alpha_{2}^{2} \sqrt{N}}\left(\cos \theta_{2}+\frac{1}{N+N^{\alpha}-1}\right) \\
    &\approx 2N^{-\frac{1}{2}} \\
    &=o(1)
\end{align*}
On the other hand, by using Eqs.~(\ref{eq:n/nn}) and~(\ref{eq:nn/n}), $k_{2}$ and $s_{2}$ are given by
\begin{equation*}
    k_{2}=
    \begin{dcases}
        -\sin\left(t\theta_{2}+\frac{3\theta_{2}}{2}\right)+o(1) &: 0 \leq \alpha <1,\\
        \frac{1}{2}\sin\left(t\theta_{}+\frac{3\theta_{2}}{2}\right)+o(1) &: \alpha=1,\\
        o(1)&: \alpha > 1,
    \end{dcases} 
    s_{2}=
    \begin{dcases}
        o(1) &: 0 \leq \alpha <1, \\
        \sin\left(t\theta_{2}+\frac{3\theta_{2}}{2}\right)+o(1) &: \alpha=1,\\
        o(1) &: \alpha >1.
    \end{dcases}
\end{equation*}
Thus, we get 
\begin{align*}
    c_{2}k_{2}=o(1), \quad c_{2}s_{2}=o(1).
\end{align*}
{\bf (v) (Estimation of $R_{K}$ and $R_{S}$):}
By using Eq.~(\ref{eq:vec_-1}), $\beta$ is given by
\begin{align*}
    \beta^{2}&=1+2\cdot (N-1) +2 \cdot \frac{(N-1)(N-2)}{N^{\alpha}} \\
    &=2N+2N^{2-\alpha}-6N^{1-\alpha}+2N^{-\alpha}-3.
\end{align*}
Hence, we have
\begin{equation*}
    \label{eq:beta_order}
    \beta^{2}=
    \begin{dcases}
        O(N^{2-\alpha}) &: 0 \leq \alpha \leq 1, \\
        O(N) &: \alpha > 1.
    \end{dcases}
\end{equation*}
By above equation, we get
\begin{equation*}
    R_{K}=
    \begin{dcases}
        O(N^{\alpha-3/2}) &:  0 \leq \alpha \leq 1, \\
        O(N^{-1/2}) &: \alpha > 1.
    \end{dcases}, \quad
     R_{S}=
     \begin{dcases}
        O(N^{\frac{\alpha-2}{2}}) &:  0 \leq \alpha \leq 1, \\
        O(N^{-\alpha/2}) &: \alpha > 1.
     \end{dcases}
\end{equation*}
Therefore we obtain $R_{K}=o(1)$ and $R_{S}=o(1)$.




\begin{thebibliography}{1}

\bibitem{ADZ93}
Y.~Aharonov, L.~Davidovich, and N.~Zagury.
\newblock Quantum random walks.
\newblock {\em Phys. Rev. A}, 48(2):1687--1690, 1993.

\bibitem{AAKV01}
D.~Aharonov, A.~Ambainis, J.~Kempe, and U.~Vazirani.
\newblock Quantum walks on graphs.
\newblock In {\em Proc. 33th STOC}, pages 50--59, New York, 2001. ACM.

\bibitem{HKSS14}
Yu.~Higuchi, N.~Konno, I.~Sato, and E.~Segawa.
\newblock Spectral and asymptotic properties of {G}rover walks on crystal
  lattices.
\newblock {\em J. Funct. Anal.}, 267(11):4197 -- 4235, 2014.

\bibitem{AKR05}
A.~Ambainis, J.~Kempe, and A.~Rivosh.
\newblock Coins make quantum walks faster.
\newblock In {\em Proc. 16th Annual ACM-SIAM Symposium on Discrete Algorithms
  SODA}, pages 1099--1108, 2005.

\bibitem{SKW03}
N.~Shenvi, J.~Kempe, and K.~B. Whaley.
\newblock A quantum random walk search algorithm.
\newblock {\em Phys. Rev. A}, 67(5):052307, 2003.

\bibitem{TSP22}
H.~Tanaka, M.~Sabri, and R.~Portugal.
\newblock Spatial search on {J}ohnson graphs by discrete-time quantum walk.
\newblock {\em J. Phys. A: Math. Theor.}, 55(25):255304, jun 2022.

\end{thebibliography}
\end{document}